\documentclass[runningheads]{llncs}
\usepackage{graphicx}
\usepackage{url}
\usepackage{multirow}
\usepackage{booktabs}
\makeatletter
\newcommand{\printfnsymbol}[1]{%
  \textsuperscript{\@fnsymbol{#1}}%
}
\makeatother
\begin{document}
\title{Heat Kernels with Functional Connectomes Reveal Atypical Energy Transport in Peripheral Subnetworks in Autism}
\titlerunning{Heat Kernels Reveal Atypical Energy Transport in Autism}
\author{Markus D. Schirmer\inst{1,2}\orcidID{0000-0001-9561-0239}\and
Ai Wern Chung\inst{3}\orcidID{0000-0002-0905-6698} }
\authorrunning{Schirmer and Chung}
\institute{Stroke Division \& Massachusetts General Hospital, J. Philip Kistler Stroke Research Center, Harvard Medical School, Boston, MA, USA \and
Department of Population Health Sciences, German Centre for Neurodegenerative Diseases (DZNE), Bonn, Germany \and
Fetal-Neonatal Neuroimaging \& Developmental Science Center, Boston Children’s Hospital, Harvard Medical School, Boston, MA, USA \email{AiWern.Chung@childrens.harvard.edu}}
%Springer Heidelberg, Tiergartenstr. 17, 69121 Heidelberg, Germany
\maketitle              % typeset the header of the contribution
\begin{abstract}
Autism is increasing in prevalence and is a neurodevelopmental disorder characterised by impairments in communication skills and social behaviour. Connectomes enable a systems-level representation of the brain with recent interests in understanding the distributed nature of higher order cognitive function using modules or subnetworks. By dividing the connectome according to a central component of the brain critical for its function (it's hub), we investigate network organisation in autism from hub through to peripheral subnetworks. We complement this analysis by extracting features of energy transport computed from heat kernels fitted with increasing time steps. This heat kernel framework is advantageous as it can capture the energy transported in all direct and indirect pathways between pair-wise regions over 'time', with features that have correspondence to small-world properties. We apply our framework to resting-state functional MRI connectomes from a large, publically available autism dataset, ABIDE. We show that energy propagating through the brain over time are different between subnetworks, and that heat kernel features significantly differ between autism and controls. Furthermore, the hub was functionally preserved and similar to controls, however, increasing statistical significance between groups was found in increasingly peripheral subnetworks. Our results support the increasing opinion of non-hub regions playing an important role in functional organisation. This work shows that analysing autism by subnetworks with the heat kernel reflects the atypical activations in peripheral regions as alterations in energy dispersion and may provide useful features towards understanding the distributed impact of this disorder on the functional connectome.

\end{abstract}
\section{Introduction}
Autism spectrum disorder is a neurodevelopmental condition estimated to affect 1 in 59 children in the US~\cite{baio_prevalence_2018}. It is characterised by atypical social behaviour and sensory processing, with deficits in high-level cognitive function and mental flexibility~\cite{hong_atypical_2019,rudie_altered_2013}. It has also been increasingly suggested that the neural bases of autism may not be explained by specific regions, but by aberrant connectivity within and between functional modules~\cite{hong_atypical_2019,muller_brain_2018}.

Thus strategies to interrogate brain connectivity in autism have evolved from specific tract analysis to connectomes. This connectome approach recognises the distributed nature of higher order cognitive functions. Recent approaches have been to identify modules or subsets of regions that are most critical for efficient network function~\cite{van_den_heuvel_rich-club_2011,schirmer_rich-club_2019,ktena_brain_2019} and which exhibit specialisation for specific processes~\cite{gollo_dwelling_2015}. For one such approach, inter-connected brain regions of high functional or structural connectivity are considered to form a collection of core “hubs”, a subnetwork that is essential for efficient cognitive function. Such hubs are the first to develop and present at birth, with strong similarity to adults~\cite{van_den_heuvel_hubs_2018,grayson_structural_2014}. Other regions that form later during development around the hubs constitute the “feeder” and “seeder” subnetworks~\cite{van_den_heuvel_high-cost_2012,schirmer_structural_2018}. Grouping connections into subnetworks allows interrogation of the relative importance of each subnetwork in characterising a disorder, such as autism, providing information on changes in the fundamental core of a network compared to secondary maladaptive changes that potentially give rise to cognitive dysfunction~\cite{van_den_heuvel_cross-disorder_2019,verhelst_impaired_2018,chung_rich_2019,chung_concussion_rich-club_2019}.  
	
Studies using subnetworks often compare measures of density or connectivity strength~\cite{verhelst_impaired_2018,collin_impaired_2013}, or traditional network metrics~\cite{schirmer_structural_2018,chung_rich_2019}. With brain function potentially being supported by coordinated activity between different functional modules, recent methods have sought to capture these dynamic processes~\cite{hong_atypical_2019,gu_controllability_2015}. Here, we propose to use heat kernels, a diffusion model, on resting-state functional MRI (rs-fMRI) data to extract features of energy transport~\cite{chung_characterising_2016,chung_classifying_2016}. The heat kernel describes the effect of applying a heat source to a network and observing the diffusion process over 'time'. It encodes the distribution of energy over a network and characterises the underlying structure of the graph~\cite{chung_spectral_1997,zhang_graph_2008}. Heat kernels have been applied to connectomes to investigate atrophy patterns in Alzheimer's~\cite{raj_network_2015}, mappings between functional and structural connectomes~\cite{abdelnour_network_2014} and for predicting motor outcome in preterm infants~\cite{chung_characterising_2016}. 
	
In this work, we present an edge-centric analysis where energy propagation features are computed from rs-fMRI hub-stratified subnetworks. These heat kernel features are then compared between a large, multi-centre cohort of autism and control subjects. By combining a dynamic network model with hub analysis, our aim is to better understand the vulnerability of and interplay between subnetworks in autism.

\section{Materials and Methods}
\subsection{Subjects and rs-fMRI data preprocessing}
We used rs-fMRI data from the Autism Brain Imaging Data Exchange (ABIDE) initiative~\cite{di_martino_autism_2014}, comprising of typically-developing controls (n=440) and subjects with autism (n=379). Data were preprocessed with the ABIDE Connectome Computation System pipeline. In brief, preprocessing steps included: removal of spikes, with slice timing and motion correction, removal of mean CSF and white matter signals, and detrending of linear and quadratic drifts. Band-pass temporal filtering (0.01-0.1Hz) was applied after the above nuisance variable regressions. The global mean signal was not regressed from the data. rs-fMRI data were registered to the MNI template and signals averaged into regions according to the AAL atlas. The pre-processed timeseries was demeaned, and a covariance matrix~\cite{varoquaux_learning_2013} was computed for each subject. Omitting brainstem and cerebellar regions resulted in a $90 \times 90$ connectivity matrix for each subject.

\subsection{Group connectomes, hub organisation and subnetworks}
\subsubsection{Group connectome}
A group-averaged network was computed from only the control population. The absolute of the connectivity matrix was taken and thresholded to retain values greater than 0.05 to remove spurious associations. A binarised, group-average adjacency matrix, $W_{group}$, was then computed by retaining edges in at least 75\% of the control group. 

\subsubsection{Hub organisation and defining subnetworks}
Hub regions were identified from $W_{group}$ by selecting the top ten nodes with the greatest strength~\cite{collin_impaired_2013}. Network edges are then labelled into subnetworks based on their connection to the hub nodes~\cite{schirmer_structural_2018}: \textit{Hub subnetwork} - Contains edges connecting two hub nodes; \textit{Feeder subnetwork} - are edges connecting hub to non-hub nodes; and \textit{Seeder subnetwork} - have edges connecting two non-hub nodes. We include a fourth \textit{'Non-edge subnetwork'}, comprising of the remaining entries in the network which do not have an actual connection. These four subnetworks form the 'regions-of-interest' to group the edge-based, heat kernel features for analysis.

\subsection{Computing heat kernels and their features}
The following computations are performed on a network W for each subject (in both control and patient groups) found by multiplying their respective covariance matrix with $W_{group}$.

\subsubsection{Graph notation}
$G=(V,E)$ where $V$ is the set of $|V|$ nodes on which a graph is defined and $E \subseteq V \times V$ the corresponding set of edges. A subject's weighted matrix, W, is defined as $W(u,v)= w_{uv}$ where $w_{uv}$ is the corresponding edge strength. A diagonal strength matrix, $D$, is defined as $D(u,u)=deg⁡(u)=\sum_{v \in V} w_{uv}$ . The Laplacian of $G$ is $\mathcal{L}=D-W$ and the normalised Laplacian is given by $\hat{\mathcal{L}} =D^{-1/2} \mathcal{L} D^{-1/2}$.

\subsubsection{Heat kernel features}
The heat kernel, $H(t)$, is the fundamental solution to the standard, partial differential equation of a diffusion process,

\begin{equation}
\frac{\partial H(t)}{\partial t} = -\hat{\mathcal{L}} H(t),
\label{eq:1}
\end{equation}

and can be computed analytically,

\begin{equation}
H(t) = \exp(-t \hat{\mathcal{L}}).
\label{eq:2}
\end{equation}

$H(t)$ describes the flow of energy through $G$ at time $t$ where the rate of flow is governed by $\hat{\mathcal{L}}$ calculated from $W$. $H(t)$ is a symmetric $|V| \ times |V|$ matrix where the entry $H_{u,v}(t)$ represents the amount of heat transfer between nodes $u$ and $v$ after time $t$.

Based on heat kernels computed from Equation~\ref{eq:2} for a range of $t$, several features can be extracted for each entry in H to represent the dynamic properties of the network~\cite{chung_rich_2019}. One measure is the intrinsic time constant, $t_c$, which is the time when the \textit{relative} change in heat transfer has dropped below a given percentage. The $t_c(u,v)$ between nodes $u$ and $v$ for percentage threshold s is

\begin{equation}
t_c(u,v) = t_{max}: \left| \frac{H_{u,v}(t + \Delta t) - H_{u,v}(t)}{H_{u,v}(t)} \right|_{t_1}^{t_2} < s,
\label{eq:3}
\end{equation}

where $\Delta t$ is a time step within the range of $t_1 \leq t \leq t_2$. The next set of measures are the maximal energy passed between two regions (maximal value across all heat kernels)

\begin{equation}
h_{peak}(u,v) = \max \left| H_{u,t}(t) \right|_{t_1}^{t_2}, 
\label{eq:4}
\end{equation}

and the time that $h_{peak}$ occurs

\begin{equation}
t_{peak}(u,v) = t: h_{peak}(u,v).
\label{eq:5}
\end{equation}

The last set of features represent the maximal \textit{difference} in energy transferred between two regions,

\begin{equation}
h'_{peak}(u,v) = \max\left|H_{u,v}(t + \Delta t) - H_{u,v}(t) \right|_{t_1}^{t_2}
\label{eq:6}
\end{equation}

and the time that $h'_{peak}$ occurs

\begin{equation}
t'_{peak}(u,v) = t: h'_{peak}(u,v).
\label{eq:7}
\end{equation}

\subsection{Experimental design}
For each subject, $1500$ heat kernels were computed from $W$ for $t=[0.00,0.01,…15.0]$. $t_c$ was calculated for a range of thresholds $s=[2,3,4,5\%]$. This results in a total of eight features for every edge in $E$. For each feature, the mean is calculated from edges belonging to each subnetwork, yielding a final $32$ measures for each subject (number of features $\times$ number of subnetworks). Group differences of these 32 measures are tested for using independent t-tests. Multiple comparisons was accounted for with a Bonferroni corrected significance threshold of $p < 0.05/32 = 0.00156$. 

\section{Results}
Table~\ref{tab:1} is an overview of subject demographics. Ages were not significantly different between groups (independent t-test,  $p=0.61$). 

\begin{table}[ht!]
  \centering
  \caption{Demographics of subjects from ABIDE}
    \begin{tabular}{ccc}
    \toprule
    Heading level & Controls & Patients \\
    \midrule
    Number of subjects & 440 & 379 \\
    Age (years, mean $\pm$ std) & 16.27 $\pm$ 6.74 & 16.53 $\pm$ 7.54 \\
    Age (years, range) & 6.47 - 56.2 & 7.0 - 58.0 \\
    \bottomrule
    \end{tabular}%
  \label{tab:1}%
\end{table}%

Regions identified as hub nodes are listed in Table~\ref{tab:2}. These regions were predominantly deep grey matter structures and have found to be key hub regions (e.g. insular, superior medial frontal, supramarginal gyrus) elsewhere~\cite{power_evidence_2013,sato_connectome_2016}.

\begin{table}[ht!]
  \centering
  \caption{Identified hub regions in controls.}
    \begin{tabular}{l}
    \toprule
    Regions \\
    \midrule
    Superior medial frontal - Left \\
    Insular - Left \\
    Insular - Right \\
    Putamen - Left \\
    Putamen - Right \\
    Supramarginal gyrus - Right \\
    Rolandic operculum - Left \\
    Rolandic operculum - Right \\
    \bottomrule
    \end{tabular}%
  \label{tab:2}%
\end{table}%

\begin{figure}[ht!]
\centering
\includegraphics[width=\linewidth]{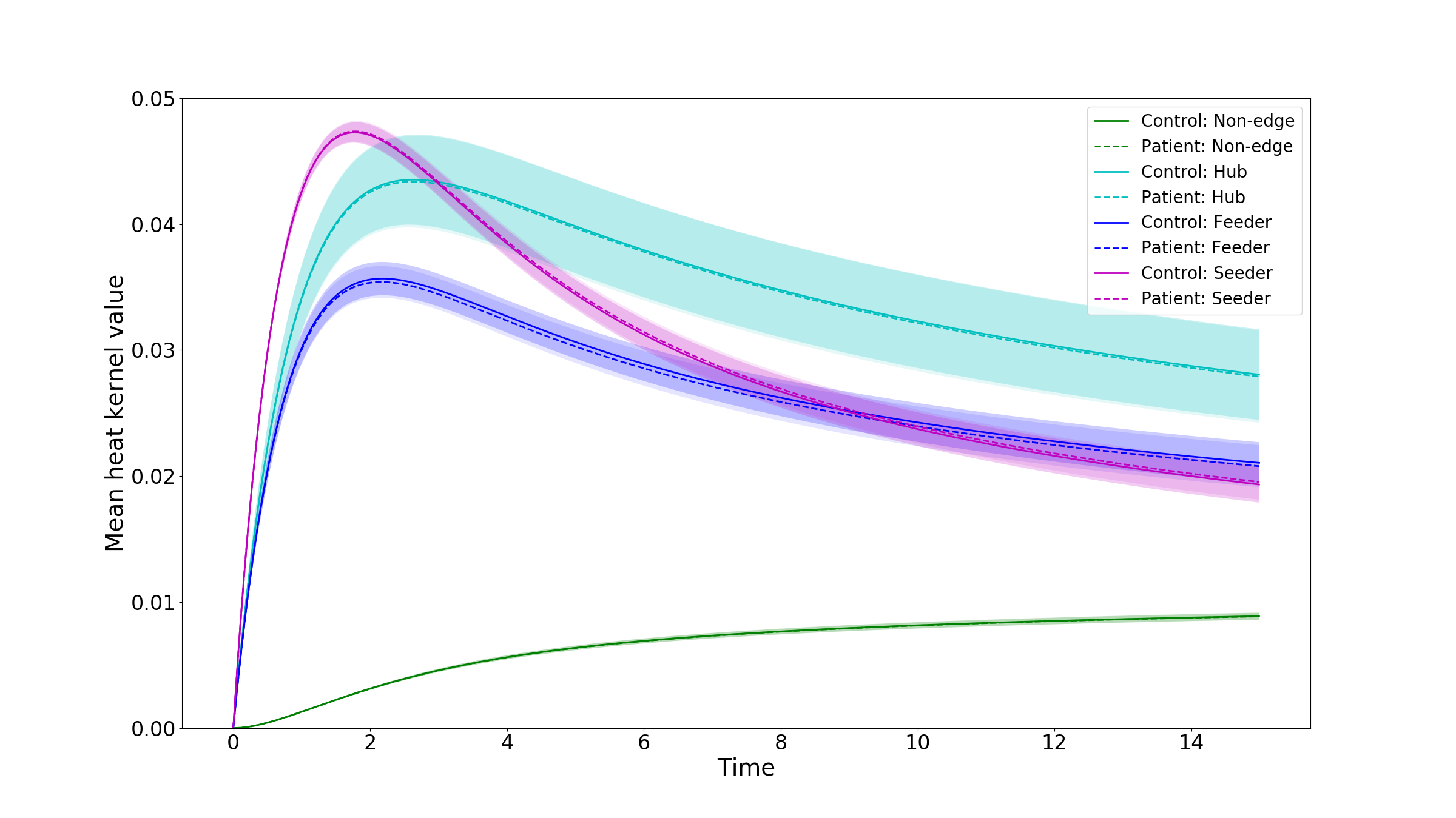}
\caption{Plots of mean values in the heat kernel matrix by subnetwork, with increasing time for all subjects in each group. Shaded areas represent standard deviations.}
\label{fig:1}
\end{figure}

Figure~\ref{fig:1} plots the amount of energy in heat kernels with time, averaged by subnetwork, for each group. Specifically, it plots the heat kernel value, $H_{u,v}(t)$, averaged across all edges within a subnetwork, versus $t$. The slope and shape of each curve varies depending on the subnetwork. The non-edge subnetwork transports the least amount of energy over time, and the remaining subnetworks all exhibit a peak where energy is maximal at different $t$. There is also a consistent difference between groups over time after the peak - patients have lower heat kernel values than controls in the hub and feeder subnetworks, and the reverse can be observed in the seeder subnetwork.

\begin{figure}[ht!]
\centering
\includegraphics[width=\linewidth]{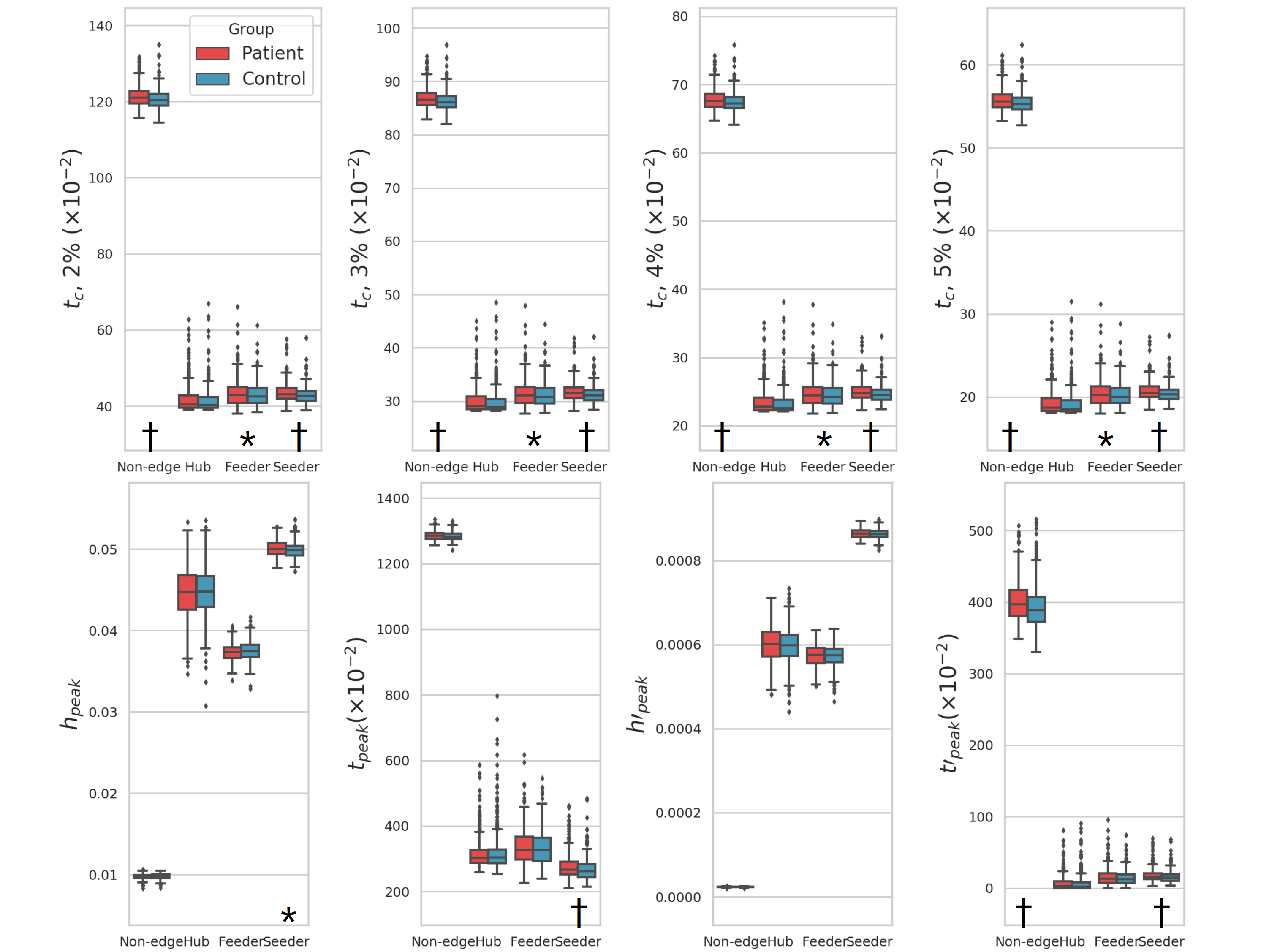}
\caption{Boxplots of mean heat kernel features by subnetwork, for controls versus autism. * denotes group differences at $p<0.05$ and $\dagger$ for statistically significant group differences corrected for multiple comparisons at $p<0.00156$.}
\label{fig:2}
\end{figure}

Figure~\ref{fig:2} plots each of the eight averaged heat kernel features, for all subnetworks and groups, and Table~\ref{tab:3} shows features' mean (standard deviation) values and statistical significance levels between groups. There are no significant group differences for all measures in the hub subnetwork, however, the more peripheral the subnetwork is to the hub, the greater the number of significant group differences, and the greater their statistical significance (Table~\ref{tab:3}). This is most apparent in $t_c$, irrespective of the threshold $s$ used. Furthermore, the autism group has greater $t_c$ than controls in all non-hub subnetworks. This trend can be similarly seen for $t_{peak}$ and $t'_{peak}$. 

\begin{table}[ht!]
\centering
\caption{Mean (stdev) heat kernel features by subnetwork for Control and autism groups. Statistically significant group differences are in bold denoted by $*p<0.05$, ${\dagger}p<0.00156$ (Bonferroni-corrected threshold), ${\ddagger}p<0.0001$.}
\label{tab:3}
\begin{tabular}{lllllllll}
    \toprule
    \multicolumn{1}{c}{\multirow{2}[2]{*}{Feature}} & \multicolumn{2}{c}{Non-edge} & \multicolumn{2}{c}{Hub} & \multicolumn{2}{c}{Feeder} & \multicolumn{2}{c}{Seeder} \\
          & \multicolumn{1}{l}{Control} & \multicolumn{1}{l}{Autism} & \multicolumn{1}{l}{Control} & \multicolumn{1}{l}{Autism} & \multicolumn{1}{l}{Control} & \multicolumn{1}{l}{Autism} & \multicolumn{1}{l}{Control} & \multicolumn{1}{l}{Autism} \\
    \toprule
    \multicolumn{1}{l}{$t_c$, 2\%} & \textbf{1.207} & \multicolumn{1}{l}{\textbf{1.215$\ddagger$}} & 0.417 & 0.418 & \textbf{0.43}  & \multicolumn{1}{l}{\textbf{0.436*}} & \textbf{0.429} & \multicolumn{1}{l}{\textbf{0.436$\dagger$}} \\
          & \textbf{(0.027)} & \textbf{(0.029)} & (0.037) & (0.035) & \textbf{(0.03)} & \textbf{(0.036)} & \textbf{0.023} & \textbf{(0.026)} \\
    \midrule
    \multicolumn{1}{l}{$t_c$, 3\%} & \textbf{0.864} & \multicolumn{1}{l}{\textbf{0.869$\ddagger$}} & 0.3   & 0.301 & \textbf{0.312} & \multicolumn{1}{l}{\textbf{0.316*}} & \textbf{0.313} & \multicolumn{1}{l}{\textbf{0.317$\dagger$}} \\
          & \textbf{(0.019)} & \textbf{(0.021)} & (0.026) & (0.025) & \textbf{(0.022)} & \textbf{(0.026)} & \textbf{(0.016)} & \textbf{(0.019)} \\
    \midrule
    \multicolumn{1}{l}{$t_c$, 4\%} & \textbf{0.674} & \multicolumn{1}{l}{\textbf{0.679$\ddagger$}} & 0.235 & 0.235 & \textbf{0.245} & \multicolumn{1}{l}{\textbf{0.248*}} & \textbf{0.247} & \multicolumn{1}{l}{\textbf{0.250$\dagger$}} \\
          & \textbf{(0.015)} & \textbf{(0.016)} & (0.021) & (0.02) & \textbf{(0.017)} & \textbf{(0.02)} & \textbf{(0.013)} & \textbf{(0.015)} \\
    \midrule
    \multicolumn{1}{l}{$t_c$, 5\%} & \textbf{0.554} & \multicolumn{1}{l}{\textbf{0.558$\ddagger$}} & 0.193 & 0.193 & \textbf{0.203} & \multicolumn{1}{l}{\textbf{0.205*}} & \textbf{0.204} & \multicolumn{1}{l}{\textbf{0.207$\dagger$}} \\
          & \textbf{(0.012)} & \textbf{(0.014)} & (0.017) & (0.017) & \textbf{(0.014)} & \textbf{(0.017)} & \textbf{(0.01)} & \textbf{(0.012)} \\
    \midrule
    \multicolumn{1}{l}{$h_{peak}$} & 0.976 & 0.974 & 4.473 & 4.465 & 3.747 & 3.733 & \textbf{4.987} & \multicolumn{1}{l}{\textbf{5.005*}} \\
          & (0.032) & (0.03) & (0.299) & (0.303) & (0.119) & (0.107) & \textbf{(0.101)} & \textbf{(0.099)} \\
    \midrule
    \multicolumn{1}{l}{$t_{peak}$} & 12.84 & 12.856 & 3.213 & 3.159 & 3.333 & 3.36  & \textbf{2.679} & \multicolumn{1}{l}{\textbf{2.760$\dagger$}} \\
          & (0.133) & (0.142) & (0.63) & (0.454) & (0.547) & (0.572) & \textbf{(0.339)} & \textbf{(0.385)} \\
    \midrule
    \multicolumn{1}{l}{$h'_{peak}$} & 0.235 & 0.235 & 5.988 & 5.985 & 5.729 & 5.733 & 8.629 & 8.641 \\
          & (0.007) & (0.007) & (0.43) & (0.424) & (0.246) & (0.248) & (0.113) & (0.108) \\
    \midrule
    \multicolumn{1}{l}{$t'_{peak}$} & \textbf{3.937} & \multicolumn{1}{l}{\textbf{4.018$\dagger$}} & 0.067 & 0.069 & 0.142 & 0.157 & \textbf{0.155} & \multicolumn{1}{l}{\textbf{0.175$\dagger$}} \\
          & \textbf{(0.294)} & \textbf{(0.306)} & (0.119) & (0.112) & (0.099) & (0.119) & \textbf{(0.079)} & \textbf{(0.091)} \\
    \bottomrule
    \end{tabular}
\end{table}

\section{Discussion}
In this work, we presented an rs-fMRI subnetwork analysis comparing features of energy propagation between autism and controls. More specifically, we investigated how heat kernel derived measures differ between groups in the central functional core of the brain and its peripheral subnetworks. We found no significant difference in energy transport in hub regions between groups. However, peripheral subnetworks differed significantly, with important properties of change in energy transport occurring at later time points in autism when compared to controls.

Combining hub-stratified subnetwork analysis with the above heat kernel framework is a complementary strategy to further our understanding of brain topology. The strategic importance of hubs for information transport makes them potentially vulnerable and thus sensitive to disease~\cite{van_den_heuvel_cross-disorder_2019,schirmer_rich-club_2019,ktena_brain_2019,crossley_hubs_2014}. Whereas alterations in the feeder and seeder subnetworks have been viewed as potential secondary adaptations to disease or injury~\cite{verhelst_impaired_2018,chung_rich_2019,chung_concussion_rich-club_2019}. However, analysis treating subnetworks as separate, standalone, entities with their own topology may be unrealistic given the highly integrative nature of the brain. Heat kernels provide a means to incorporate information from the entire network, even when computing edge-wise measures. This is because each element in $H$ represents energy transport through all possible pathways that connect any two regions. It is because of this that the non-edge subnetwork possesses heat information, and has the lowest heat kernel value. The non-edge subnetwork is also highly indicative of a network's capacity (its small world propensity) for efficient energy propagation when using heat kernels~\cite{chung_characterising_2016}, explaining the greatest significance between groups of all subnetworks tested in our analysis.

This combined framework revealed a number of interesting findings. The lack of significant group differences in the hub subnetwork suggests a relatively preserved functional core in autism. Subnetwork analysis in other pathologies have identified core regions to remain similar to controls, whereas peripheral regions demonstrated greater differences~\cite{verhelst_impaired_2018,chung_rich_2019}. While it has been suggested that the core is stable in order to support and allow peripheral regions greater flexibility to meet functional demands~\cite{gollo_dwelling_2015}, others have found atypical functional activation in the core in the ABIDE cohort against controls~\cite{hong_atypical_2019,keown_network_2017}. Differences in reported hub connectivity in autism may be attributed to not only the different methodologies used, but also because of the many ways to rate a node's importance to be labelled as a hub. 
 
In terms of peripheral subnetworks, the recruitment of more seeder regions has been found in long, indirect functional pathways in autism than in controls~\cite{hong_atypical_2019}. The authors suggest this may be indicative of diminished segregation between core and peripheral subnetworks. It is thus interesting that the more peripheral the subnetwork, the greater the statistical significance of our group differences measured by heat kernel features. This result in non-hub regions was also accompanied by greater values in autism than in controls for all time related features. Suggesting that while heat kernel values have a similar profile in both subject groups  (Figure~\ref{fig:1}), the extracted features quantifying properties of these curves appear at a later t in autism. As it stands, we cannot ascertain what these changes in the heat kernel features and their timings represent, but taken all together, our results suggest greater involvement from peripheral, rather than hub regions in autism.

There are limitations to our study, one of which is the broad age range in the ABIDE dataset. It is important to understand the potential impact of age on our results, particularly as the age range encompasses a period of great neurodevelopmental change from childhood through to early adulthood. Another is our use of node strength to identify hubs when a measure which includes information on shortest path lengths such as betweenness centrality may be more relevant, even though there is great overlap between hubs found using both metrics on ABIDE data~\cite{keown_network_2017}. Future work will take these limitations into consideration and incorporate other methods to determine nodal importance~\cite{van_den_heuvel_rich-club_2011,schirmer_network_2018}.

In this study, we present a novel analysis by combining two complementary frameworks of energy propagation and subnetworks to investigate differences in network efficiency in a large autism and control dataset. Overall, we identify significant group differences in all peripheral subnetworks (feeder, seeder, non-edge) and a preserved central hub in the autism group, further supporting the key role that non-centralised regions play in brain functional organisation. How these energy transport features in the peripheral subnetwork are related to cognitive function and their association with clinical measures in autism remain to be determined.

\subsection*{Acknowledgments}
This project has received funding from the American Heart Association and Children’s Heart Foundation Postdoctoral Fellowship, 19POST34380005 (AWC) and the European Union’s Horizon 2020 research and innovation programme under the Marie Sklodowska-Curie grant agreement No 753896 (MDS).

\bibliographystyle{../splncs04}

\end{document}